\documentclass[prb,aps,onecolumn,showpacs,floats,a4paper,times,12pt]{revtex4}
\usepackage{epsfig,bm,epsf,graphics}
\usepackage{xcolor}
\begin{document}
\draft
\title{Sociophysics of Intractable Conflicts: Three-Group Dynamics}
\author{Miron Kaufman$^{a}$ \footnote{ Corresponding author, E-mail:m.kaufman@csuohio.edu},  Hung. T. Diep$^b$ and Sanda Kaufman $^c$  }

\address{
$^a$ Department of Physics, Cleveland State University, Cleveland, OH 44115, USA\\
$^b$ Laboratoire de Physique Th\'eorique et Mod\'elisation,
Universit\'e de Cergy-Pontoise, CNRS, UMR 8089\\
2, Avenue Adolphe Chauvin, 95302 Cergy-Pontoise Cedex, France\\
$^c$ Levin College of Urban Affairs, Cleveland State University, Cleveland, OH 44115, USA\\
}

\begin{abstract}
We extend a sociophysics model of two-group conflict dynamics to three groups. The model with attractors and chaos is proposed as a tool for exploring and managing intractable conflicts. It can be used to generate scenarios of trajectories and outcomes. We use mean-field theory for long-range interactions to study the time dependence of the three grousp' mean attitudes. We find that at some intermediate temperatures the group mean attitudes  oscillate in time.  Independent of initial conditions, trajectories converge overtime to an attractor in the three-dimensional space of mean attitudes. We use Monte Carlo simulations to explore short-range group interactions and find chaotic unpredictable time variation of attitudes at high temperatures. For illustrative purposes we apply the model to the Bosnia and Herzegovina conflict.
\end{abstract}

\maketitle

\section{Intractable social conflicts and sociophysics}

Social conflicts have been the focus of negotiation and conflict management  research for several decades. \cite{foot1}  Groups in conflict differ not only in their interests but also in identity, values and beliefs (e.g., see Ref. \onlinecite{Oberschall}), and conflict frames. When lasting for relatively long time periods that sometimes exceed the lifespan of the generation that started the strife, these conflicts are deemed intractable, resistant to resolution, or in Valacher et al.'s  terms\cite{Valacher},"seemingly intractable."\cite{foot2} Bercovitch \cite{Bercovitch}  observed that intractable conflicts self-perpetuate, are difficult to manage, and revolve not only around differences of the moment but also around deep-rooted issues; some give rise to violent episodes. Burgess and Burgess (2003)\cite{Burgess} classified the causes of intractability into irreconcilable value or moral differences, high-stakes distributional issues (resources) and/or domination issues (power). Intractable conflicts often have a mix of causes from all three categories.

Depending on the context and scale of a conflict and on the types of potential negative consequences, management tools include diplomacy, mediation, advocacy on behalf of a stakeholder group, legal procedures, attempts to foster peaceful dialog between disputing groups, and even military interventions. Media play a (not always constructive) role in such processes \cite{Menczer}. Besides knowledge of the stakeholders' interests, their history, cultures, interests, institutional arrangements and other specific case-based details, managing conflicts requires ability to explore possible futures in order to construct effective strategies. Intractable conflicts are challenging in this respect: they are complex and emergent, and embedded in interrelated and changing, broader social systems. Therefore, predicting outcomes based on the situation of the moment is likely to be misleading.
However, to strategize, make a move, or to accept/reject proposals for settling a conflict, groups need to evaluate possible results of alternative courses of action. Anticipatory scenarios may be helpful in exploring various possible futures \cite{Bernstein,Diep2017,Kaufman1,Kaufman2,Kaufman3}.

A useful set of tools for generating anticipatory scenarios in multiparty, complex, intractable conflicts comes from sociophysics. Several authors, for instance Wilson (1969), Stauffer (2003), Galam (2012), Barnes and Wilson (2014), Godoy et al. (2017), and Schweitzer (2018),\cite{Wilson,Stauffer,Galam1,Barnes,Godoy,Schweitzer} offer overviews of how physics tools have been applied to various aspects of social processes. Examples include attitude changes \cite{Galam2}, economics \cite{Stanley}, social network measures \cite{Liben}, social dynamics \cite{Castellano}, and social interaction processes \cite{Helbing}. Modeling of various aspects of intractable conflicts is illustrated by Coleman et al. (2007)\cite{Coleman1}, Liebovitch et al. (2008)\cite{Liebovitch}, and Kaufman and Kaufman (2013)\cite{Kaufman1}.

Sociophysics is not without critics. Majorana and Mantegna (2006) \cite{Majorana} question the degree of similarity between physical particles and people, who unlike particles, have agency. After deploring that "God gave physics the easy problems," Bernstein et al. (2000)\cite{Bernstein} argued that the analogy between physical and social phenomena is weak, if judged by the poor predictive performance of social models built as if they were representing physical phenomena. Their critique refers chiefly to the lack of precise, broadly accepted definitions for social variables of interest, unlike in physics, as well as lack of sufficient numbers of similar cases to afford generalizations and reliable predictions. Indeed, although both physical and social systems are complex and display emergent behaviors, unlike physical particles humans do not follow predictable rules, such that given the same conditions the same results obtain. Bernstein et al. (2000)\cite{Bernstein} described how molecules differ from people for prediction purposes. Their critique is apt for attempts, as Stewart (1950)\cite{Stewart} had proposed, to predict social behaviors using physics methods despite these differences.

Sociophysics models, however, do not necessarily seek prediction in the same sense as social models. Rather, by using physics analogies, they help capture patterns and trajectories of complex social systems, yielding a range of future possibilities that decision makers in a particular situation need to take into account as they prepare strategies.
While  discouraging the use of physics models for social theory building and making point predictions, Bernstein et al.'s (2000) statement that \cite{Bernstein}, "Knowledge of structure and process also allows conscious and far-reaching transformations of social systems" supports the use of such models in efforts to understand interactive systems in order to intervene and change course when necessary.

Statistical physics models of social systems with a large number of members, each interacting with a subset of others, have been used in very diverse domains such as culture dynamics, crowd behavior, information dissemination  \cite{Castellano} and social conflicts \cite{Diep2017, Kaufman3, Coleman}. Buchanan (2007) \cite{Buchanan} observed that such models rely on the fact that large societal groups display surprising regularities  despite individual agency. In response to the critique of oversimplification of social dynamics when representing them with statistical physics tools, Castellano et al. (2000) \cite{Castellano} note that "in most situations qualitative (and sometimes quantitative) properties of large scale phenomena do not depend on the microscopic details of the process. Only higher level features, such as symmetries, dimensionality or conservation laws, are relevant for the global behavior." \cite{foot3} Thus, it is not necessary to assume that humans behave mechanistically, which is a critique frequently leveled at sociophysics models.

In this article, we illustrate the utility of sociophysics to the study of intractable social  conflicts by considering the political situation in Bosnia and Herzegovina (BiH), the site of long-standing conflicts (e.g., Friedman 2013 \cite{Friedman}). In the next section we extend to three groups the Diep et al. (2017) model\cite{Diep2017} of two groups in conflict.  In Section III, we describe the three  groups in the BiH conflict, which has flared up again ahead of  2018 presidential elections. In Sections IV and V we report  results  of our model applied to the BiH situation, using the mean-field approach for long-range interactions, and Monte Carlo simulations for short-range interactions. Section VI is a qualitative discussion of the model results in general and for the BiH conflict.In Section VII we offer concluding remarks.

\section{The Model}\label{model}

We assume, as in the two-group model\cite{Diep2017}, that in each of three disputing groups each individual has a preference or attitude $s$, such that  $-M_n \leq s \leq M_n$ (n=1, 2, 3), regarding whether or not to engage with the other group to resolve the conflict. Individuals whose attitudes $s = -M_n$ or close to it are the most open to compromise, being only very loosely attached to their group's identity or ideology. Those with attitudes around $M_n$ stick to their goals to the point of confrontation if necessary, due to strong adherence to their own group's ideology, and willingness to defend it by any means. This attitude leads to opposition to concessions. The 0 midpoint of this range represents adherence to the values of one's group, combined with willingness to find a way out of the conflict with the opposing groups.

As in the two-group model, we also assume that each group is a network of members interacting with each other. This linkage pattern among members of a group based on some shared characteristics is called homophily (see McPherson et al.\cite{McPherson} and Aiello et al.\cite{Aiello}). In the words of McPherson et al., "similarity breeds connection". The networks can interact with each other, forming a multiplex. Within each group, each individual acts with a certain intensity to persuade others to his/her point of view, and is in turn subject to others' persuasion efforts. In any group, the individuals' stances are also affected indirectly by the "average" stances of the other groups.

The three-group model yields group preference averages $s_n (n=1, 2, 3)$ at any time t, which result from the members' intra-group mutual interactions, and consideration of the average attitudes of opposing groups. The intra-group intensity of advocacy (which we conceptualize as negative energy) of an individual from group $n$ is $J_n*s*s_n$, where $s_n$ is the average of all individual preferences in group $n$. When an individual's stance is affected by another group, the inter-group intensity of interaction (negative energy) is taken to be proportional to the product between that individual's preference s and the mean value of the preferences of the other groups' members. For example, for an individual in group 1: $K_{12}*s*s_2 + K_{13}*s*s_3$, where $K_{1j}(j=2,3)$ captures the individual's reaction to the average attitudes in groups 2 and 3.

The two-group model\cite{Diep2017} had 4 parameters (2 $J_s$ and 2 $K_s$); for three-group there are 9 parameters (3 $J_s$ and 6 $K_s$), and in general an $n-$group model will have $n^2$ parameters ($n$ $J_s$ and  $n^2-n$ $K_s$), rapidly increasing the levels of computational and representation difficulty. However, useful insights can be derived from the $n=3$ model.

Unlike in physics \cite{Diep2015}, the matrix of inter-group interactions is not necessarily symmetrical: $K_{mn} \neq K_{nm}$ because of human agency. While physics phenomena obey Newton's third law,the magnitudes of human action and reaction do not have to be equal. Rather, the effect of group $n$ on group $m$ can be different in magnitude and sign from the effect group m has on group n. When $K_{mn}$ and $K_{nm}$ have opposite signs  a "frustration-like" effect emerges: a  positive $s_{m}$ induces a negative $s$ value in group $n$ because of negativity of $K_{nm}$; a negative $s_{n}$ induces negative $s$ value  in group $m$ because of positivity of $K_{mn}$.Thus while group $n$ acts on group $m$ positively, eliciting a "tit-for-tat" response in the latter (similar values of the corresponding attitudes), group $m$ may act on group $n$ negatively, eliciting "contrarian" responses.\cite{foot4}

A temperature, reflecting contextual factors and quantified in the  model by means of the Boltzmann probability distribution, drives the variability in individual preferences $s$ in a group. Our dynamic model captures the evolution of group preferences by assuming that the intensity of interactions involves the product of individuals' preferences at a current time and average preferences of opposing groups at an earlier time. This lag reasonably reflects the fact that results of individuals' persuasion efforts in one time period are likely to materialize in a later time period.

As in the two-group model, we study the three-group multiplex first using a mean-field approach and then with Monte Carlo (MC) simulations, because each approach captures different aspects of the interactions. We illustrate next the kinds of insights this model can provide by applying it to the Bosnia and Herzegovina conflict.

\section{Example: Bosnia and Herzegovina}\label{BH}

Numerous three-group conflicts at different scales, intra-national as well as international are intractable, or protracted. They can be latent over stretches of time, and occasionally erupt in violent episodes that mark people's memories for a long time. Specific events can trigger the flare-ups. Both international and intra-national conflicts can have ethnic, political, resources and environmental issues. They share characteristics captured by our model: they are resistant to resolution and unfold over long time periods, recurring even after having been seemingly settled for a while; the three disputing groups are homophilic and differ from each other in values, immediate and long-term interests, in the intensity with which they involve themselves in the disputes, in their internal cohesion - the strength of linkages between group members - and in the types and strength of their  linkages with the other groups. The three groups are relatively stable, and therefore identifiable over time, although some of their characteristics such as group size vary. The Bosnia and Herzegovina (BiH) case shares these traits.

BiH's  Bosniaks, Serbs, and Croats are headed to presidential elections in 2018 with a heightened level of ethnic strife (Tamkin 2018\cite{Tamkin} and Sito-Sucic 2018\cite{Sito-Sucic}) that otherwise ebbs and flows along the years. Some basic data convey a sense of scale. With 20 square miles (51km2) BiH had a population of approximately 3.5 million at the last census of 2013, published in 2016 \cite{Toe} and contested by Serbs. Half the population is Bosniak (Muslim). The other half is composed of Serbs (Orthodox, 31\%) and Croats (Catholic, 15\%). Formerly part of Yugoslavia, BiH declared independence (and was recognized internationally) in 1992. That year, a violent interethnic war erupted, which ended with the General Framework Agreement for Peace in Bosnia and Herzegovina (the Dayton Accords) of 1995 between the presidents of Bosnia, Serbia and Croatia.

The Dayton Accords stipulated BiH's complicated political structure \cite{Mertus,Friedman}. It consists of a federation of Bosniaks and Croats (FB\&H, on 51\% of the territory) and the Republika Srpska (RS, on 49\% of the territory). The small, multiethnic Br\u{c}ko District (population 83,000) was created in 2000 on land from the other two districts. Each of the three ethnic groups, namely Bosniaks, Serbs and Croats, elects a president, \cite{Szasz} with Br\u{c}ko allowed to vote with either the FB\&H or RS. The chairship of this presidency, with an 8-months term, rotates among the three elected presidents.

A survey of attitudes among the BiH population (Prism 2013\cite{Prism}) brings out some of the issues currently fueling conflicts both between the three ethnic groups and with the government (see a summary of some of the issues in Table 1 ). Among the biggest reasons for dissatisfaction, 72\% cited corruption, followed by the economy and the politics (59 and 51\% respectively) and the government's performance (27\%). Only 10\% of respondents said that relations with people from another ethnic group were problematic. Perhaps this is because BiH is highly segregated: most respondents reported living in areas where their ethnic group is dominant (with few wishing it were different), so there may be few encounters with people from the other groups.The percentage of respondents proud of their ethnic identity (93\%) roughly matched the percentage with pride in religious identity, and 88\% were proud of their regional identity (the three identities greatly overlap) while only 73\% felt proud of being BiH citizens. Serbs felt least attached to BiH (at 47\%) and Bosniaks most (at 91\%). This indicates the three groups are homophilic and the union is not stable yet.

The respondents' prediction of a new armed conflict within 5 years has been correct: only one third thought it very or somewhat likely, with the Serbs least pessimistic (42\% thought armed conflict unlikely by 2018). Should BiH break apart, only 38\% expected it to happen peacefully, with 59\% skeptical of this possibility. Half of respondents also wished to separate their (national/ethnic) territory from BiH. With their own Republika Srpska, Serbs were least interested in this possibility and half of them were disengaged, while Bosniaks and Croats expressed more readinesss to take arms to defend their territory.

BiH respondents mostly concurred that the government, politicians and the international community were most responsible for their problems. These contextual factors driving  dissatisfaction correspond to temperature in our model. The respondents holding the context repsonsible  may account for their relatively high reported levels of apathy; only about 50\% are prepared to participate in the elections, with Bosniaks at the lowest level (46\%) and Croats highest at 66\%. Slightly more than a third of the population would like to leave BiH. Most predicted no change in the political arrangement among the three groups, but if they could have their wishes only about one third of FB\&H and only a quarter of Serbs would keep this arrangement. Half of the Serb respondents preferred an independent RS. Almost no one expressed interest in Serbs joining Greater Serbia or Croats joining Greater Croatia. 40\% of respondents (with 60\% of Bosniaks) felt that eventually the ethnic entities FB\&H and RS should be abolished in favor of one Bosnia.

The underlying latent ethnic conflict is ever-present despite the seeming apathy. About 30\% of respondents believe they need to reach consensus around a common truth about the past in order to move forward. However only about 13\% think they should discuss their past grievances, or believe reconciliation is possible or important; only 12\% agree that they should ignore past grievances and focus on the future. Reconciliation is not for tomorrow: only 40\% believed in 2013 that it would happen in 5 to 10 years, with Croats most optimistic at 63\% and Serbs most pessimistic at 28\%.\cite{foot5}

The flare-up around elections, as well as previous unrests (e.g. 2014), are part of a long stream of manifestations of the underlying long-term intractable ethnic/religious conflict which resurfaces - at times violently - around general discontent or events such as the breakup of the former Yugoslavia or the presidential election of 2018. For example, in the Prism survey\cite{Prism}, few (17\%) are members of a political party to advance their views, while 76\% of respondents were disengaged to angry (76\% of Bosniaks, 66\% of Croats, and 70\% of Serbs) with low expectations of improvement.

To study the BiH conflict dynamics, we use the Prism survey data\cite{Prism} to guestimate the values of the 9 parameters of the three-group  model, and then explore qualitatively (as with a toy model) the pattern of possible outcomes of the conflict. In what follows, Bosniaks are group 1; Serbs group 2 and Croats group 3.

The parameter values were chosen to reflect the three groups' homophily. Serbs have the highest internal cohesion (intragroup interaction) followed by Croats and Bosniaks: $J_2 > J_3> J_1$. The inter-group interactions are positive, or "Tit-for-Tat"\cite{foot6}, except for Bosniaks who have been militarily dominated by Serbs and Croats. Hence if the other groups engage in hostile behavior, Bosniaks may respond in a conciliatory manner to avert violence; however, if the other two groups are rather conciliatory, Bosniaks may find the space to defend their interests more aggressively. To capture these kinds of reactions we assume that $K_{12} < 0$, $K_{13} < 0$, meaning "non Tit-for-Tat"  Bosniak responses to Serbs and Croats. Note this instance of a social situation's departure from physical systems' symmetry of $K$ values, due to human agency. In terms of magnitude, inter-group interactions $K$ are ordered as follows: Bosniaks-Serbs $>$ Croats-Bosniaks $>$ Serbs-Croats, since larger groups should have stronger interactions.


\begin{widetext}
\begin{center}
\begin{table}
\begin{tabular}{|l|c|c|c|c|r|}
\hline
Issue $\Downarrow$ \ \ Percentage (\%) $\Rightarrow$ & B\&H (Total) & Bosniak & Serb & Croat & Other  \\
\hline
&&&&&\\
Voting at elections &  52.4 & 46.4     &    66.4    & 60.1 & 31.0    \\
Joining a political party & 17.1 & 13.2 &  23.6   &   22.1   &  12.4     \\
Joining a citizens' action group & 16.4 &  14.9 & 22.5 &  17.4 & 12.1    \\
Taking action such as demonstrating &  26.6  &  26.9  & 40.6   &  21.2  & 30.0    \\
Using violence or force if it becomes necessary  &  6.5  & 5.5 & 11.8   &  6.5  & 3.7  \\
Leave BiH &  36.3  & 32.2 & 46.0   &  38.3  & 56.2  \\
&&&&&\\
\hline
\end{tabular}
\caption{ Answers by ethnicity to the question: To what extent are you willing to take part in the following activities? (Prism 2013 \cite{Prism}) \label{table} }
\end{table}
\end{center}
\end{widetext}

We show next mean-field and Monte Carlo simulation approaces results for the BiH case.

\section{Mean-Field Approach}\label{MF}
\subsection{Equations}

The Renyi-Erd\"{\o}s network corresponds to our homophily assumptions about how the individuals in the three BiH groups interact with each other. The mean of preferences s of each group is proportional to the exponential of the intensity of interactions (negative energy):

\begin{eqnarray}\label{eq1}
s_1(t+1)&=&\frac{\sum_{s=-M_1}^{M_1}se^{s(j_1<s_1>(t)+k_{12}<s_2>(t)+k_{13}<s_3>(t))} }{\sum_{s=-M_1}^{M_1}e^{s(j_1<s_1>(t)+k_{12}<s_2>(t)+k_{13}<s_3>(t))}},\label{eq1a}\\
s_2(t+1)&=&\frac{\sum_{s=-M_2}^{M_2}se^{s(j_2<s_2>(t)+k_{23}<s_3>(t)+k_{21}<s_1>(t))} }{\sum_{s=-M_2}^{M_2}e^{s(j_2<s_2>(t)+k_{23}<s_3>(t)+k_{21}<s_1>(t))}},\label{eq1b}\\
s_3(t+1)&=&\frac{\sum_{s=-M_3}^{M_3}se^{s(j_3<s_3>(t)+k_{31}<s_1>(t)+k_{32}<s_2>(t))} }{\sum_{s=-M_3}^{M_3}e^{s(j_3<s_3>(t)+k_{31}<s_1>(t)+k_{32}<s_2>(t))}}\label{eq1c}
\end{eqnarray}

where $j_n=J_n/T$ and $k_{n,m} = K_{n,m}/T$ for $n$,$m$ = 1, 2 , 3. We use units such that $k_B = 1$. 		
We introduce the lag time in equations (1) by letting the preference $s$ at time $t + 1$ interacts with the averages $s_1$, $s_2$ and $s_3$ evaluated at an earlier time $t$. Here time is measured in units of the delay time. The sums on the right hand sides of equations (1)-(3) involve the Brillouin function:

\begin{equation}\label{eq2}
B(x,y,z,j,k,l,m)=(M+\frac{1}{2})\coth[(M+\frac{1}{2})(jx+ky+lz)]-\frac{1}{2} \coth[\frac{1}{2}(jx+ky+lz)]
\end{equation}

Equations \ref{eq1a}-\ref{eq1c} can be written as:
\begin{eqnarray}
s_1(t+1)&=&B(s_1(t),s_2(t),s_3(t),j_1,k_{12},k_{13},M_1)\label{eq3a}	\\
s_2(t+1)&=&B(s_2(t),s_3(t),s_1(t),j_2,k_{23},k_{21},M_2)\label{eq3b}	\\
s_3(t+1)&=&B(s_3(t),s_1(t),s_2(t),j_3,k_{31},k_{32},M_3)\label{eq3c}			
\end{eqnarray}

This model is driven by three values $M$ (number of states for each individual in 3 networks), by 3 values of the intra-network couplings $J$, and 6 values of the inter-network couplings $K$. We consider next only $M_1 = M_2 = M_3 = 3$. To completely describe the qualitative behavior of the time dependence of $s_1$, $s_2$, and $s_3$ we explore a 9-dimensional parameter space. Model results for BiH are shown in subsection IVB. Figure 1 is an example that is qualitatively different in dynamics from the BiH, illustrating the model's versatility.







\begin{figure}[h!]
\centering
\includegraphics[scale=0.40]{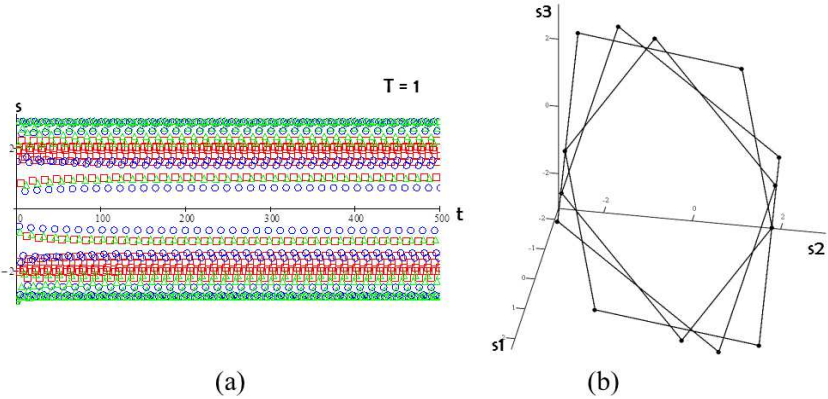}
\caption{(a) Time oscillations of group attitudes between a finite number of values. In order not to clutter the figure, the points are not connected by lines ; (b) Emerging attractor over time (see text for comments). In this figure and for all figures, groups 1, 2 and 3 are represented by red, blue and green symbols, respectively. $J_1 = 0.10$, $J_2 = 0.25$, $J_3 = 0.15$, $K_{12} = - 0.10$, $K_{21} = 0.10$, $K_{23} = -0.50$, $K_{32} = 0.50$, $K_{31} = 0.20$, $K_{13} = -0.20$ and $T = 1.0$. }
\label{ffig1}
\end{figure}

For $J_1 = 0.10$, $J_2 = 0.25$, $J_3 = 0.15$, $K_{12} = - 0.10$, $K_{21} = 0.10$, $K_{23} = -0.50$, $K_{32} = 0.50$, $K_{31} = 0.20$, $K_{13} = -0.20$ and $T = 1.0$ there are oscillations among a discrete number of values. Each group attitude has several different values represented by the same color in Fig. \ref{ffig1} (group 1 red, group 2 blue, group 3 green). The emerging attractor has a set of 14 points that are sampled over time and connected by lines in chronological order (Figure 1b). If the temperature is increased or lowered away from $T=1.0$ the attractor becomes a continuous curve.







\vspace{1cm}

\subsection{Bosnia \& Herzegovina Results}

We show next the qualitative dependence of the attitude dynamics for different temperatures, with mean-field results for $J_1 = 0.15$, $J_2 = 0.35$, $J_3 = 0.25$, $K_{12} = -0.2$, $K_{21} = 0.2$, $K_{23} = 0.1$, $K_{32} = 0.1$, $K_{31} = 0.15$, $K_{13} = -0.15$. These values are selected to represent the BiH situation (as discussed in Section III).

\begin{figure}[h!]
\centering
\includegraphics[scale=0.40]{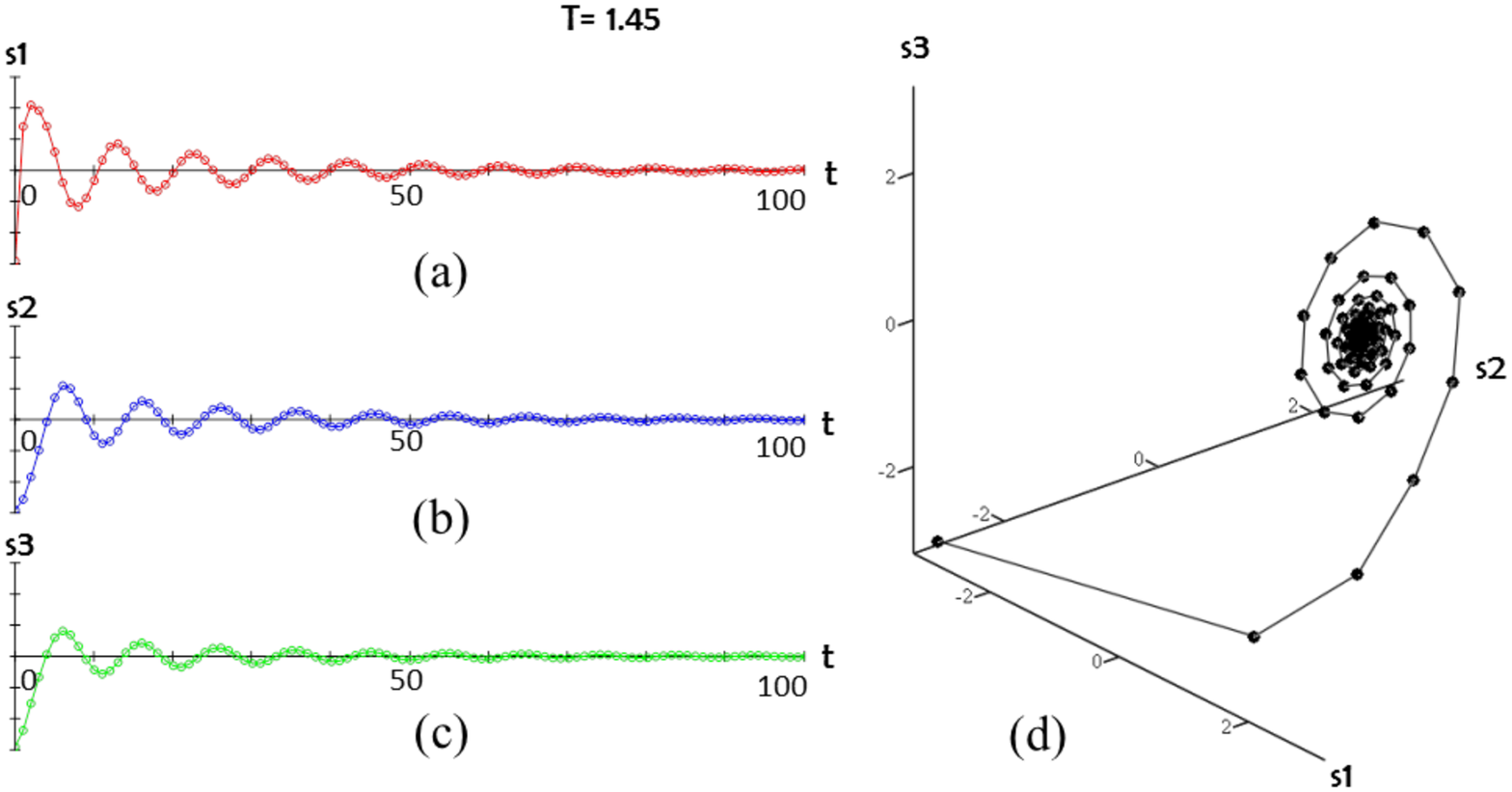}
\caption{ (a)-(c) Damped oscillations at high $T$ for groups 1, 2 and 3, respectively: $T=1.45$; (d) Spiral trajectory  to disorder. For color codes, see Fig. 1. $J_1 = 0.15$, $J_2 = 0.35$, $J_3 = 0.25$, $K_{12} = -0.2$, $K_{21} = 0.2$, $K_{23} = 0.1$, $K_{32} = 0.1$, $K_{31} = 0.15$, $K_{13} = -0.15$, selected to represent the BiH situation}
\label{ffig2}
\end{figure}

At high temperatures, representing a context that exacerbates the BiH situation, such as upcoming elections or international events affecting internal BiH affairs, the three groups' attitudes $s$ converge to zero over time  (see Fig. \ref{ffig2}), with damped oscillations as  function of time.  In the three-dimensional space $(s_1,s_2,s_3)$ a spiral trajectory in time converges to the origin,$s_1 = s_2 = s_3 = 0$.  This is a disorder point where, for each group, all possible attitudes are equally probable, resulting in the zero average.

\begin{figure}[h!]
\centering
\includegraphics[scale=0.40]{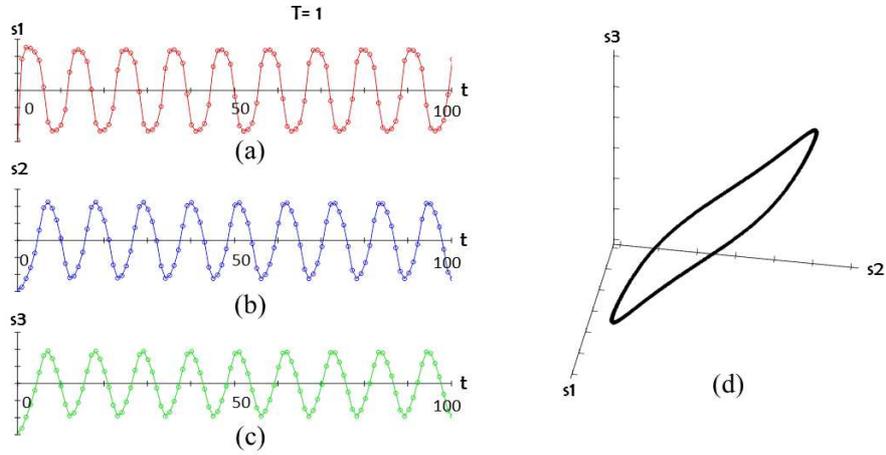}
\caption{(a)-(c) Synchronized sustained oscillations for groups 1, 2 and 3, respectively at $T=1.0$, (d) Closed loop attractor. For color codes, see Fig. 1. Values of parameters: see Fig. 2.}
\label{ffig3}
\end{figure}

At lower temperatures $T = 1.0$ the oscillations are sustained and the three groups' attitudes are synchronized exhibiting the same period.(see Fig. \ref{ffig3}). The trajectory in the $(s_1, s_2, s_3)$ space evolves in the long run to a closed-loop attractor shown in Fig. \ref{ffig3}.

\begin{figure}[h!]
\centering
\includegraphics[scale=0.40]{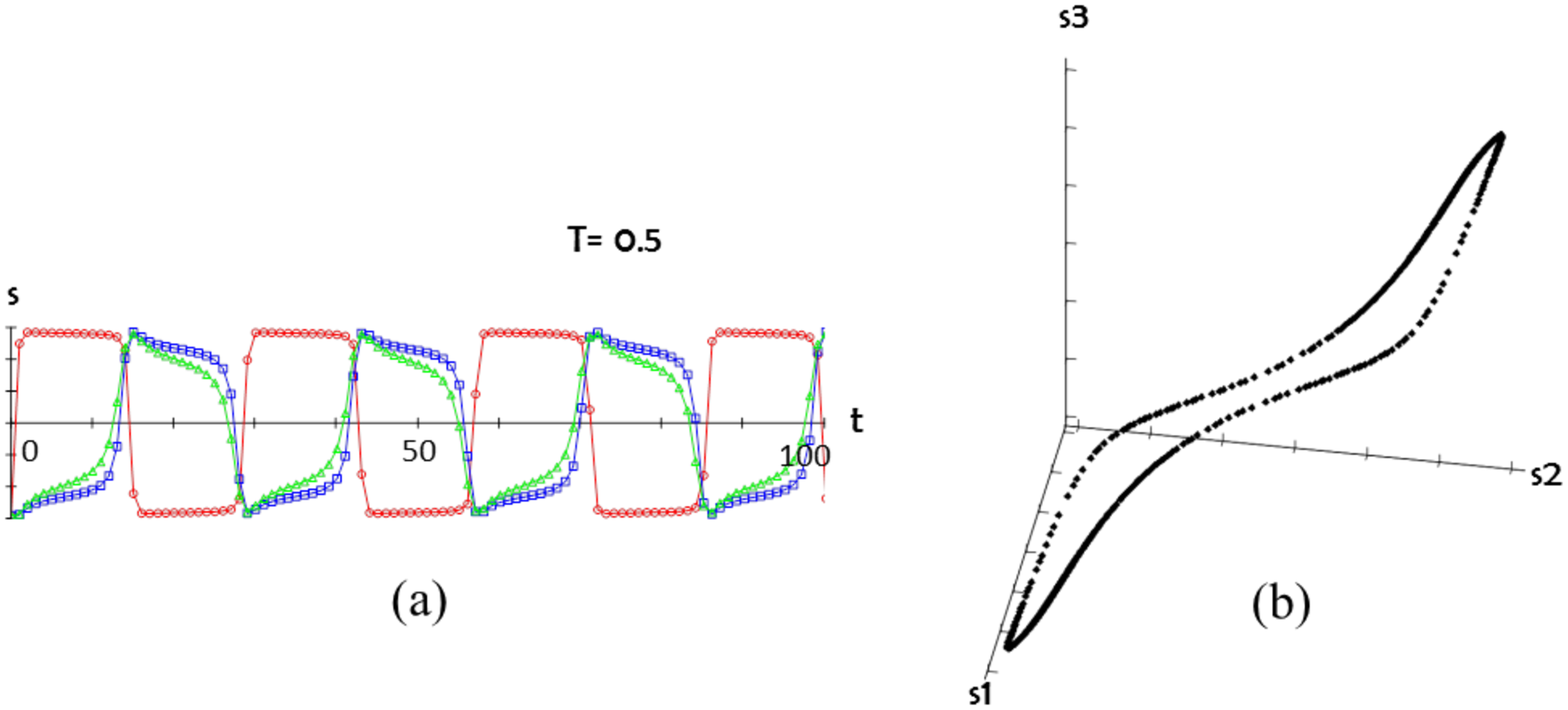}
\caption{(a) Synchronization, (b) Fragmented attractor. $T=0.5$. For color codes, see Fig. 1.  Values of parameters: see Fig. 2.}
\label{ffig4}
\end{figure}

Lowering the temperatures increases the period of oscillations, and the closed-loop attractor begins to fragment (Fig. \ref{ffig4} and Fig. \ref{ffig5}).

We argue that the higher temperature attractor represents the essence of acute intractability: there is no single point at which the conflict settles, but rather a never-ending (non-sequential) cycling occurs among possible outcomes on the attractor.

For (context-driven) lower temperatures the attractor fragments into a discreet number of fixed points. The system still cycles among them but the discrete configuration corresponds to a lower degree of intractability than the continuous attractor (Fig. \ref{ffig2}):

\begin{figure}[h!]
\centering
\includegraphics[scale=0.40]{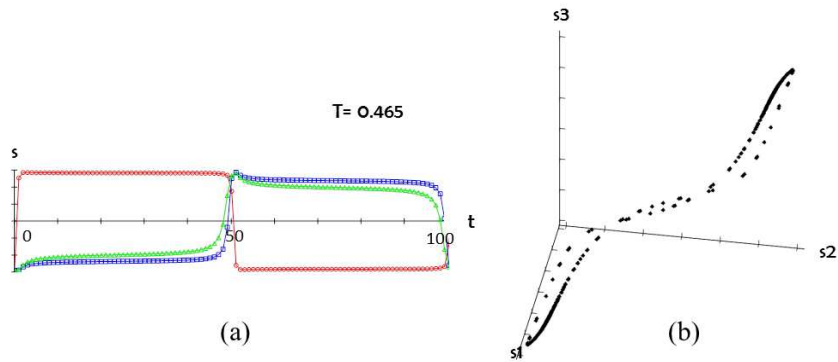}
\caption{(a) Synchronization, (b) Fragmented attractor. $T=0.465$. For color codes, see Fig. 1.  Values of parameters: see Fig. 2.}
\label{ffig5}
\end{figure}

\begin{figure}[h!]
\centering
\includegraphics[scale=0.40]{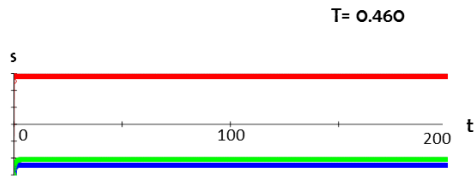}
\caption{Fixed point, corresponding to static model ordered phase. $T=0.460$. For color codes, see Fig. 1.  Values of parameters: see Fig. 2.}
\label{ffig6}
\end{figure}

At even lower temperatures the attractor collapsse into a single point that corresponds to the ordered phase of the static model (Fig. \ref{ffig6}). Note that because of the up-down symmetry of the model $S \rightarrow -S$ there is another long term solution where all the attitude values are replaced by their negatives.

\section {Model for Monte Carlo simulation}\label{MCsection}

In this section, we use the same model as the mean-field (MF) model above, but we take into account only short-range intra-group interactions. Each individual in group $n$ has a finite number of neighbors with whom he/she interacts. The intra-group Hamiltonian of group $n$ is
\begin{equation}
H_{n}=-J_{n}\sum_{i,j}s_i s_j
\end{equation}
where the sum is taken over nearest neighbors (NN) $i$ and $j$ belonging to group $n$, with interaction $J_n$.
For each group we use a triangular lattice network: each individual occupies a site having six NN with whom he/she interacts. The use of the triangular lattice allows us, when necessary, to introduce a percentage of intra-group frustration to take into account ''rebel'' attitudes of some individuals. We will come back to this point in the discussion.

We assume hereafter that all individuals have the same interaction $J_i$ with their neighbors. The strength of $J_i (i=1,2,3)$ gives each-group stability against perturbations from outside such as the social temperature in which the group is immersed, and the influence of the other groups.

For canonical MC simulations, we express the temperature $T$ from the MF model (Diep et al. 2017)\cite{Diep2017} as follows
\begin{eqnarray}
j_1 (MF)&=& \frac{J_1}{k_BT}, \ \ \ j_2 (MF)= \frac{J_2}{k_BT} \ \ \ j_3 (MF)= \frac{J_3}{k_BT} \label{mc1}\\
k_{12} (MF)&=& \frac{K_{12}}{k_BT}, \ \ \ k_{13} (MF)=\frac{K_{13}}{k_BT} \ \ \ k_{21} (MF)=\frac{K_{21}}{k_BT}, \ \ \ etc. \label{mc2}
\end{eqnarray}
where $k_B$ is the Boltzmann constant to be set to 1 hereafter.

The simulation is carried out as follows: at a given set of $J$ parameter values, we start by equilibrating each group without inter-group interaction. This is done with the standard Metropolis algorithm: for each individual we calculate the field acting on him/her by the NN. This yields an energy $E_1$. The individual chooses his/her personal state to have the new energy $E_2$ with a probability proportional to $\exp[-(E_2-E_1)/T]$. We see that if $E_2>>E_1$ then the probability of taking the state $E_2$ is almost zero. Conversely, if $E_2\leq E_1$ the change of his/her state always occurs. We achieve a MC step $t$ by doing this for all individuals of the group. We repeat the steps numerous times until the equilibrium is reached (time-independence of physical quantities).

\begin{figure}[h!]
\centering
\includegraphics[scale=0.40,angle=0]{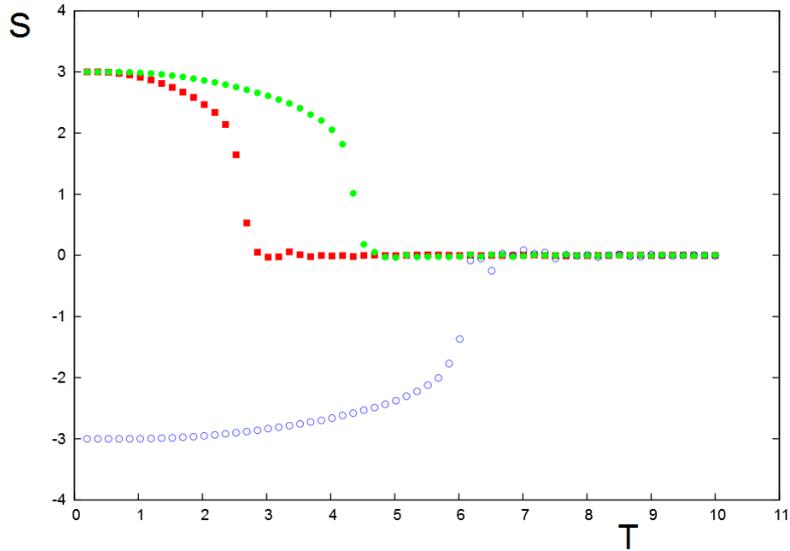}
\caption{Stability of the 3 groups as a function of social temperature $T$ before inter-group interactions are turned on. $J_1 =0.15$,  $J_2 =0.35$,  $J_3 =0.25$.  For color codes, see Fig. 1.}
\label{ffigMC1}
\end{figure}

An example is shown in Fig. \ref{ffigMC1}. We see that the social temperature $T_C$,  beyond which the stability of a group is lost, is not the same for three groups: the higher the intra-group interaction $J$, the higher $T_C$ threshold.


As in the MF calculation above, an individual in a given group interacts at the time $t+1$ with the average of the action field created by the other groups at the earlier time $t$. The only difference from the MF calculation is the short-range interaction considered in the MC simulation.  We will see that the results differ in some important aspects.

Once the equilibrium is reached for each group, we turn on the interactions between groups at time $t$. As described in section \ref{model}, we calculate at time $t$ the normalized "force" field of group $J_n$  acting on the other groups as
\begin{equation}
h_n(t)=\sum_{i\in group\ n} s_i(t)/N_n
\end{equation}
where $N_n$ is the population of group $n$.  An individual from group $m$ at the time $t+1$ interacts with his/her NN and with the group $n$ at the earlier time $t$ by
\begin{equation}
-K_{mn}s_m(t+1)h_n(t)
\end{equation}
where $m\neq n$. We calculate the strengths $s_n(n=1,2,3)$ as functions of $t$, at a given $T$.

An example at low $T$ is shown in Fig. \ref{ffigMC1} where the inter-group interactions may or not destroy the order of a group.
We have chosen the interaction strengths and signs in the example below to illustrate the BiH case described in section \ref{BH}.


\begin{figure}[h!]
\centering
\includegraphics[scale=0.25,angle=0]{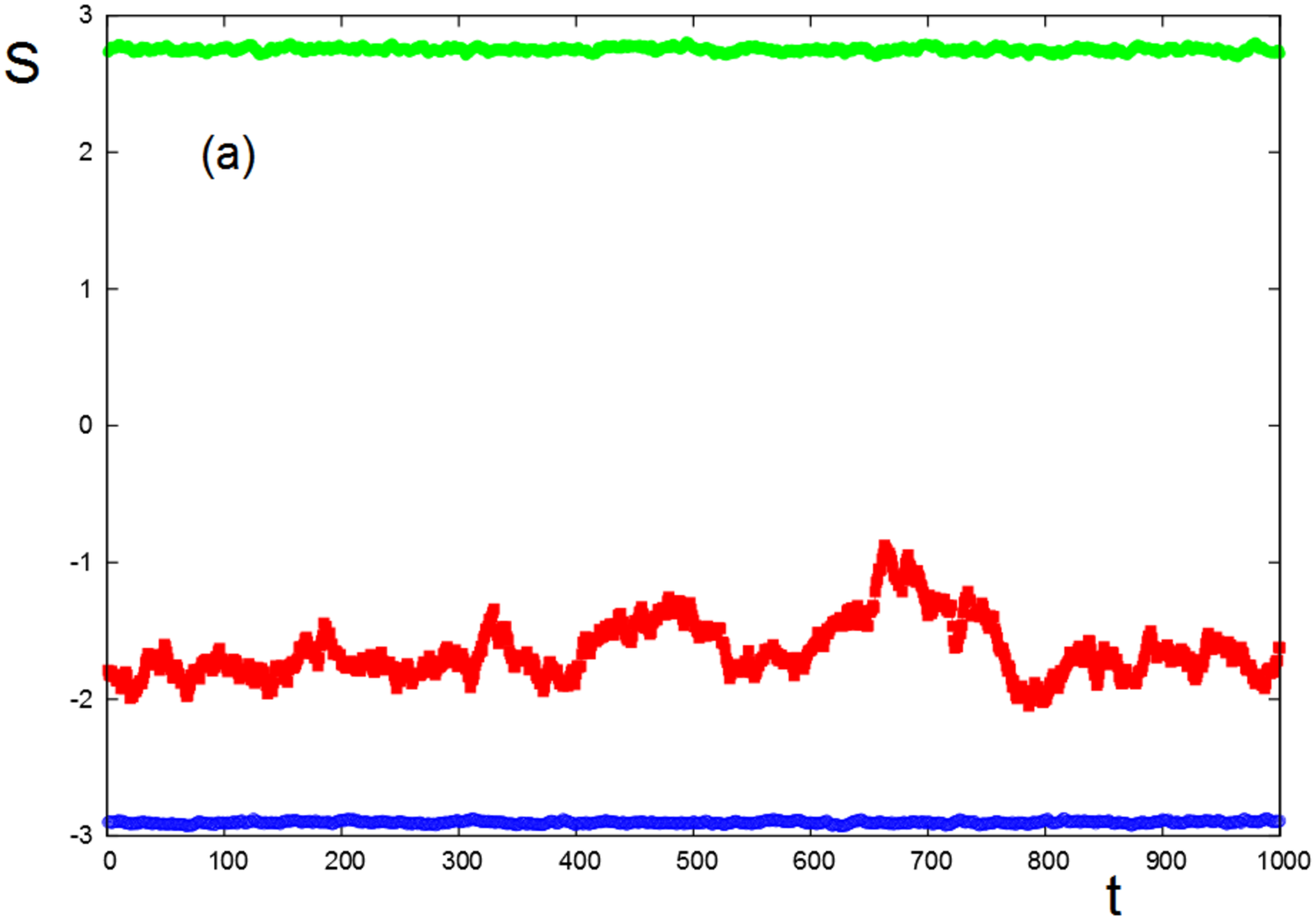}
\includegraphics[scale=0.25,angle=0]{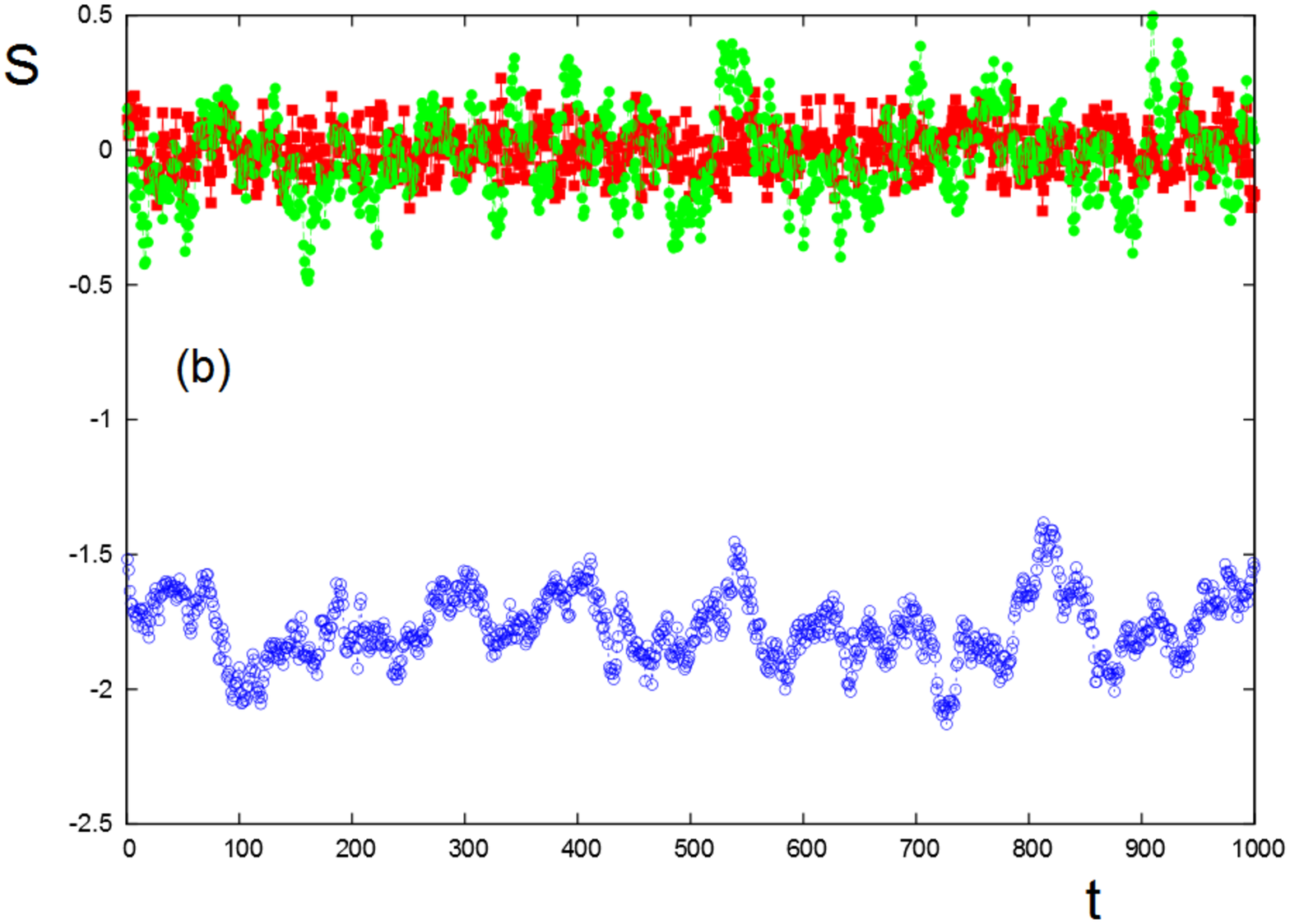}
\includegraphics[scale=0.25,angle=0]{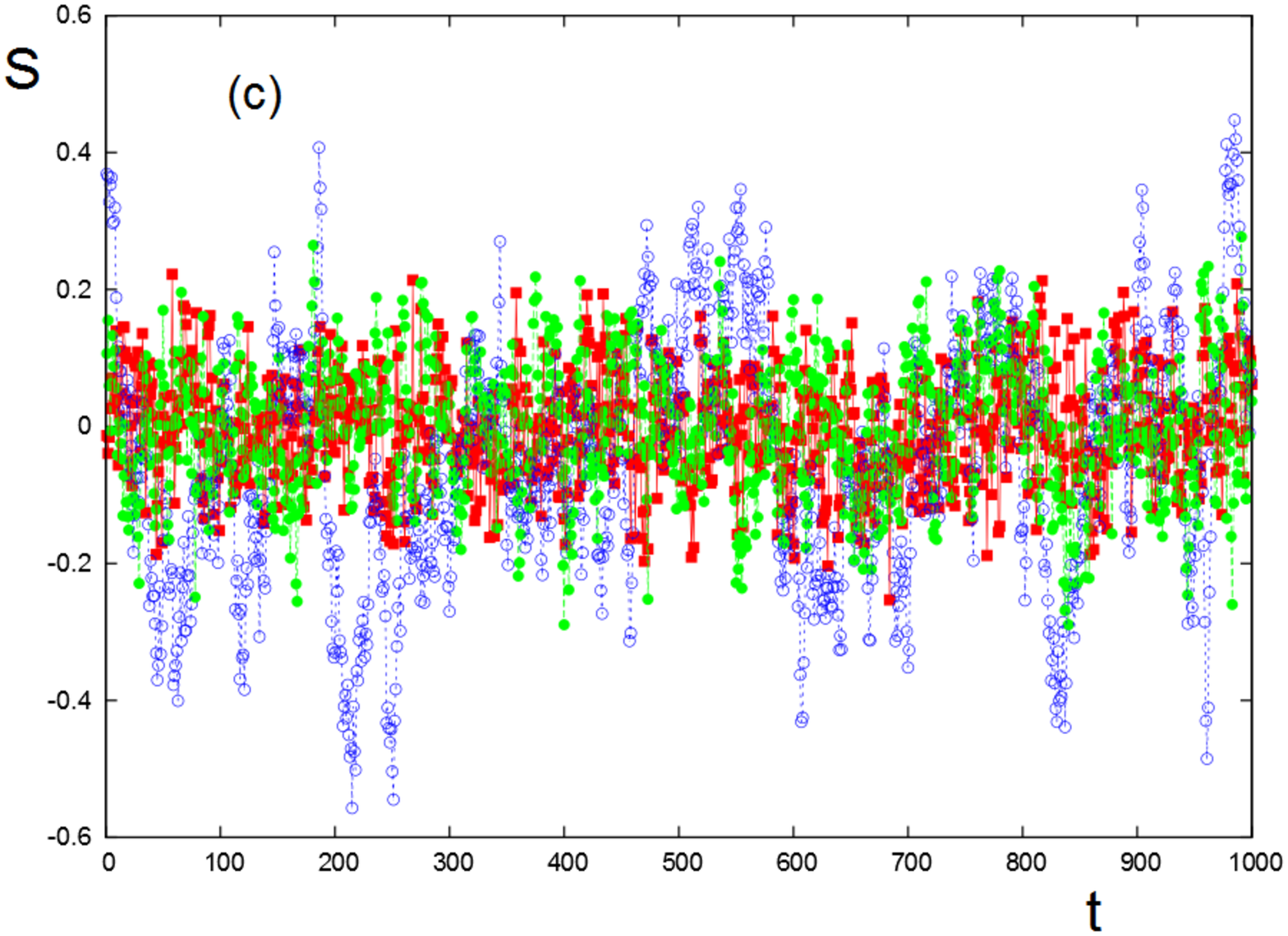}
\caption{Time-dependence of 3 groups' stances at low temperatures (for color codes, see Fig. 1): (a) $T=2.5254$,all three groups are ordered; (b) $T=5.8474$, groups 1 and 3 are disordered, group 2 is not disordered; (c) $T= 7.5084$, all 3 groups are disordered. The same parameters as in Fig. \ref{ffigMC1} have been used: $J_1 =0.15$,  $J_2 =0.35$,  $J_3 =0.25$, $K_{12} = - 0.20$, $K_{21} = 0.20$, $K_{13} = -0.15$, $K_{31} = 0.15$, $K_{23} = 0.10$, $K_{32} = 0.10$.}
\label{ffigMC2}
\end{figure}

At higher $T$, the order of each group is weakened. The inter-group interactions cause the groups' stances to oscillate widely without periodicity as also seen in the long-range MF results above. We observe that at times the stronger group 2 dominates the other two. This pattern reflects the level of intractability of the three-group conflict simulated here, consistent with the longer-term MF results. While the conflict is intractable at all the temperatures of Figure \ref{ffigMC2}, at the lower temperature (corresponding to a stable context) the groups are `stuck' in predictable ways (see Fig. \ref{ffigMC2}a); as the context gets heated, the three-group system cycles unpredictably through various stages (Fig. \ref{ffigMC2}b and c).

%

%

\begin{figure}[h!]
\centering
\includegraphics[scale=0.25,angle=0]{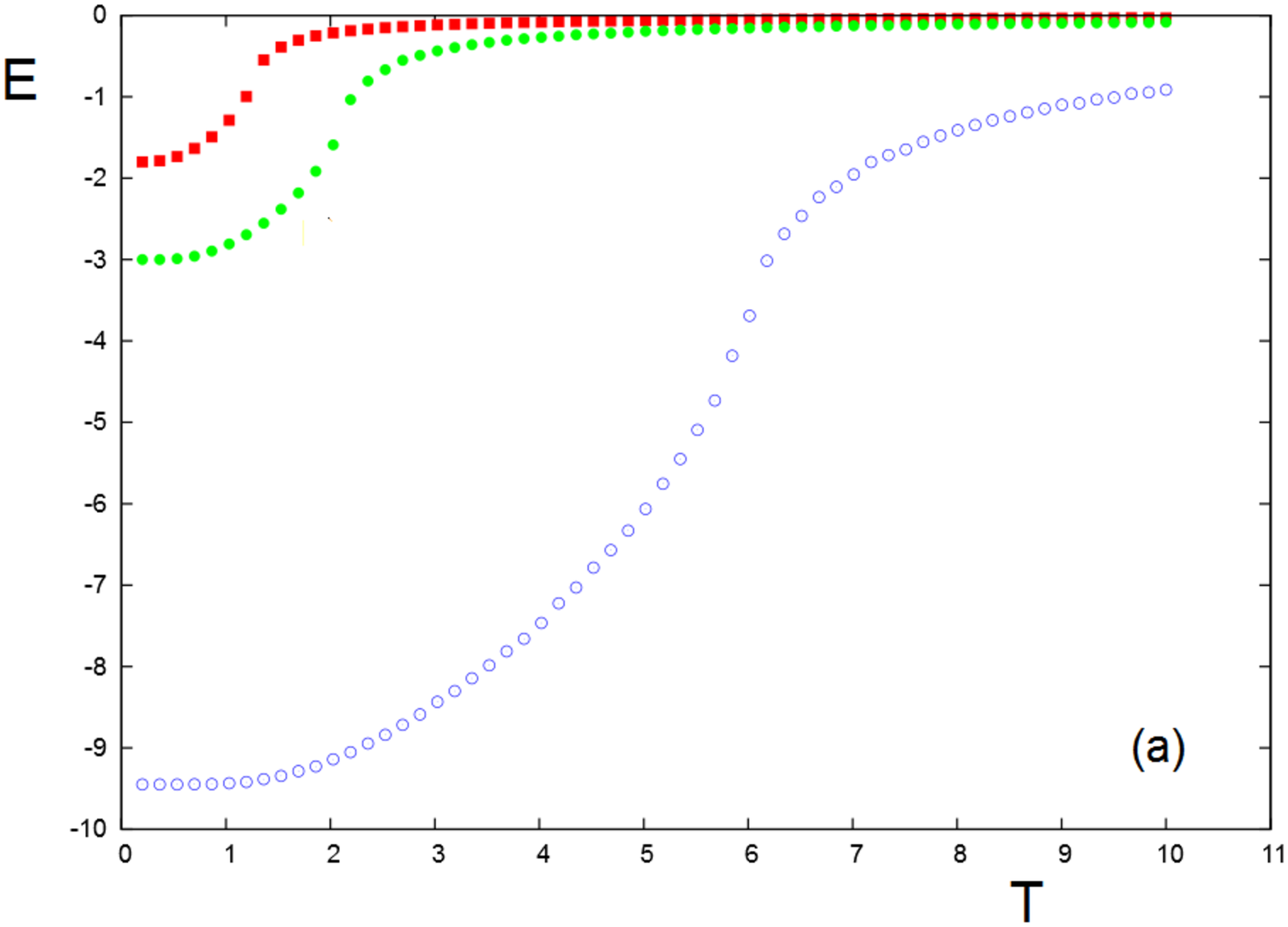}
\includegraphics[scale=0.25,angle=0]{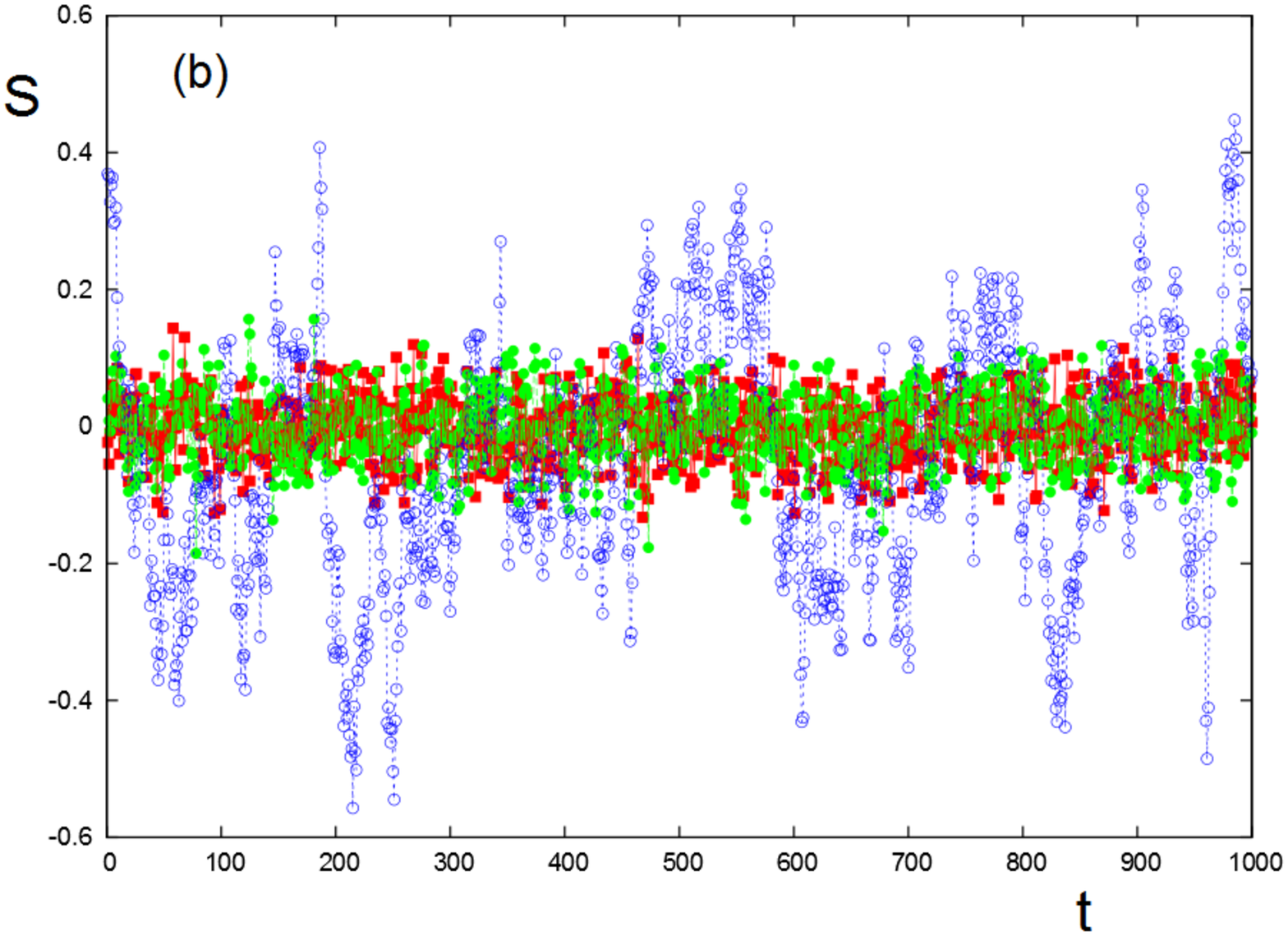}
\includegraphics[scale=0.25,angle=0]{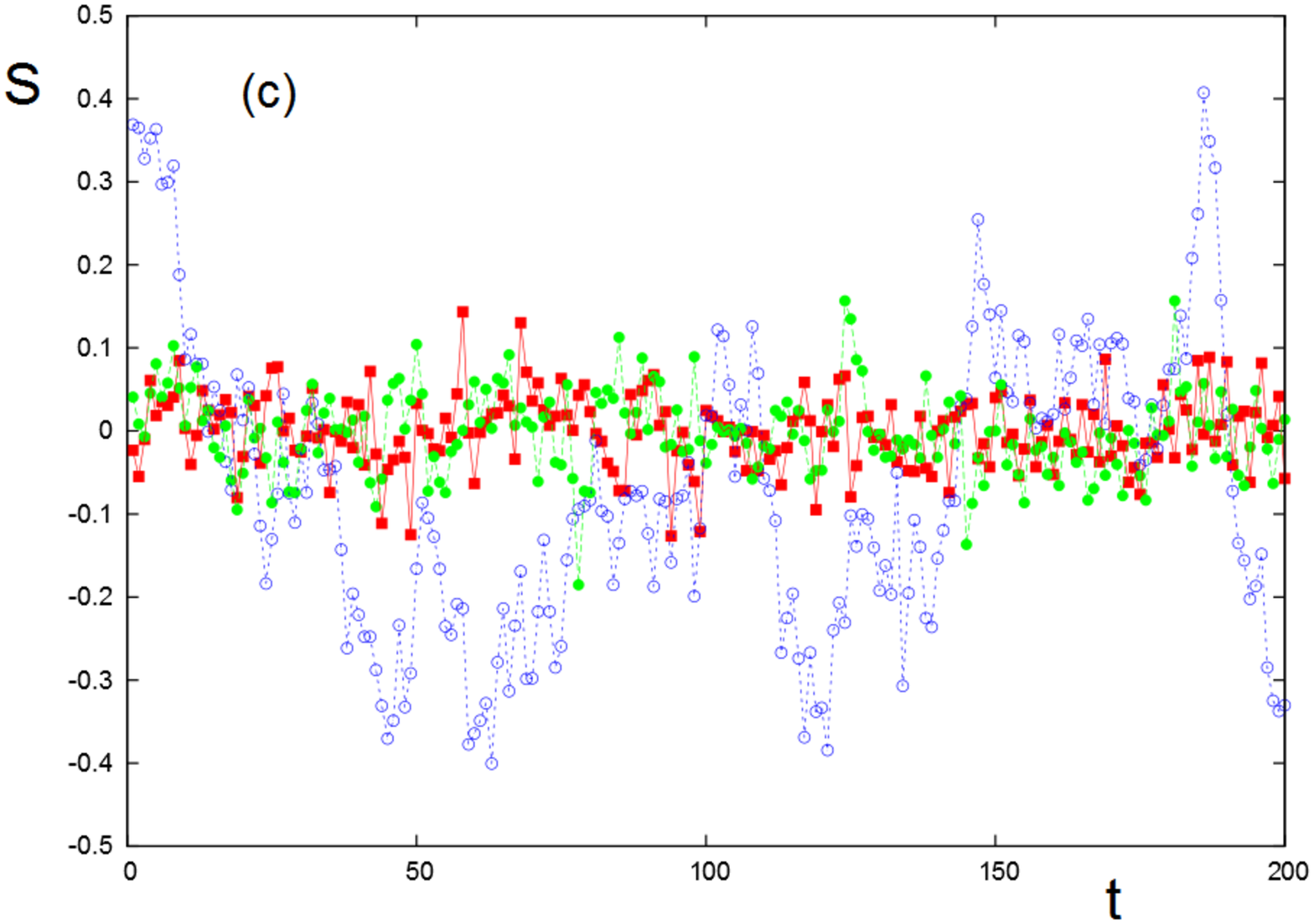}
\caption{Effect of number of states $Q_1=2M_1+1=5$, $Q_2=2M_2+1=7$, $Q_3=2M_3+1=5$, $M_1=2$, $M_2=3$, $M_3=2$ (for color codes, see Fig. 1) (a) Zoom of the time-dependence of 3 groups' stances at high $T$: $T=7.5084$; (b) Time-dependence of the strengths of 3 groups in interaction (c) Zoom of the oscillation. The same parameters as in Fig. \ref{ffigMC1} have been used: $J_1 =0.15$,  $J_2 =0.35$,  $J_3 =0.25$, $K_{12} = - 0.20$, $K_{21} = 0.20$, $K_{13} = -0.15$, $K_{31} = 0.15$, $K_{23} = 0.10$, $K_{32} = 0.10$.}
\label{ffigMC5}
\end{figure}

Let us show briefly in Fig. \ref{ffigMC5} the results when the number of intra-group individual states differs among groups.  The diminution of $Q$ causes a weaker intrinsic strength of the group as seen by the internal energies of Group 1 and 3 in Fig. \ref{ffigMC5}a.  At high $T$ Groups 1 and 3 exhibit similar random oscillation, while the strong Group 2's domination  is very pronounced as seen in Fig. \ref{ffigMC5}b and c.

We have also examined the effect $J_i$.  Some differences in details are found but the overall random oscillation when 3 groups are at high $T$ remains.

\begin{figure}[h!]
\centering
\includegraphics[scale=0.25,angle=0]{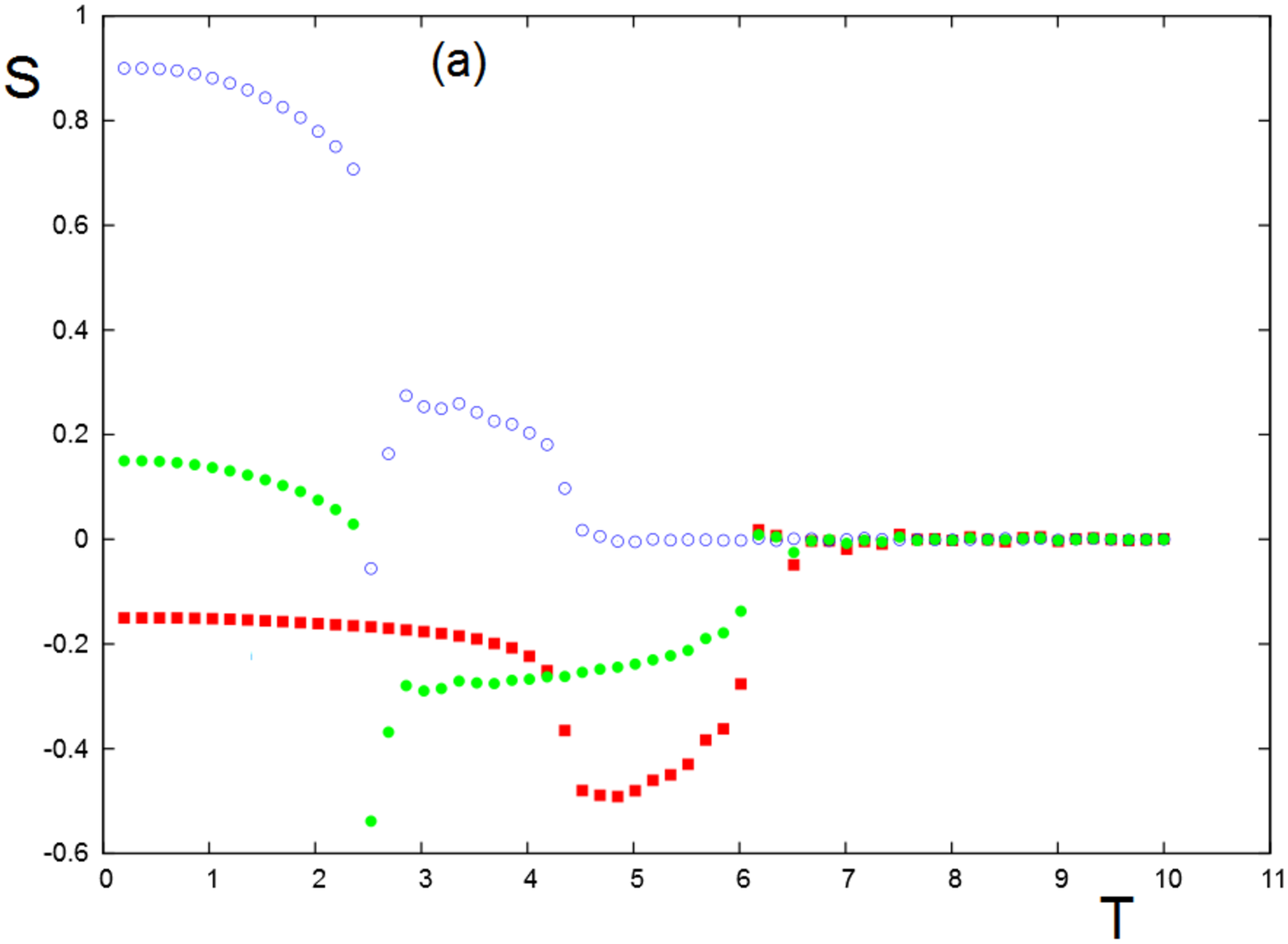}
\includegraphics[scale=0.25,angle=0]{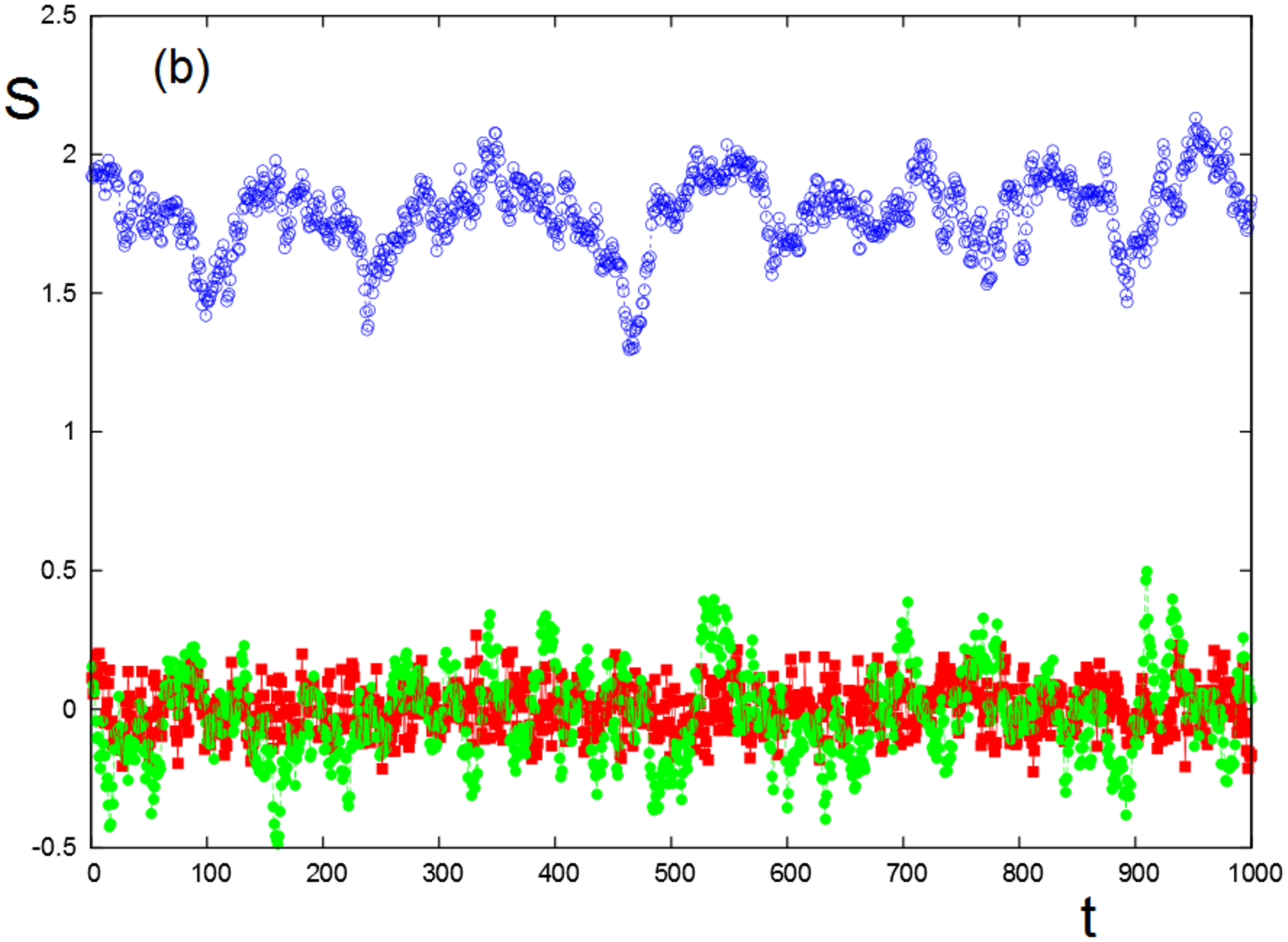}
\includegraphics[scale=0.25,angle=0]{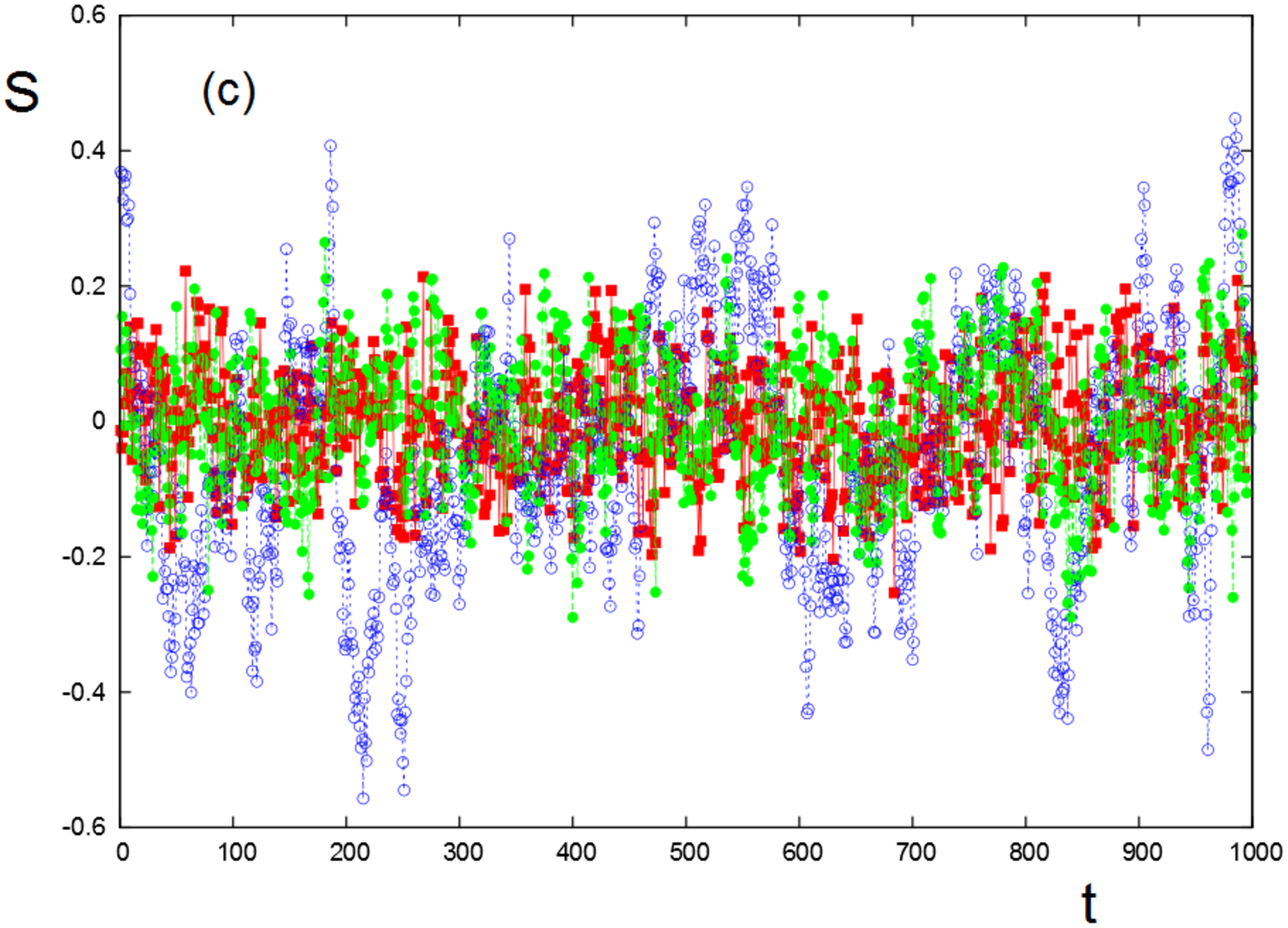}
\caption{Effect of $K_{ij}$: $Q_1=Q_2=Q_3=7$, $M_1=M_2=M_3=3$ (for color codes, see Fig. 1) (a) Long-time average of group strengths as function of $T$ (b) Time-dependence of 3 groups' stances at $T=5.8474$ (c) Time-dependence of 3 groups' stances at high $T$: $T=7.5084$. The same parameters as in Fig. \ref{ffigMC1} have been used: $J_1 =0.15$,  $J_2 =0.35$,  $J_3 =0.25$, but $K_{13}$ and $K_{32}$ change their sign: $K_{12} = - 0.20$, $K_{21} = 0.20$, $K_{13} = +0.15$, $K_{31} = 0.15$, $K_{23} = 0.10$, $K_{32} = - 0.10$.}
\label{ffigMC7}
\end{figure}

%
The effect of $K_{ij}$ has been investigated. Within small changes with respect to the values of $K_{ij}$ used in Fig. \ref{ffigMC1} do not change qualitatively the results. However, when the signs of $K_{ij}$ change, there are interesting changes in the long-time averages of the group strengths as seen in Fig. \ref{ffigMC7}a.  The short-time oscillating behavior on the other hand does not change.

\section{Discussion}\label{discussion}

Our discussion of BiH conflict specifics and the insights for it derived with mean-field computations and Monte Carlo simulations is combined with discussion of some of the three-group model's general features and implications for applying sociophysics to the analysis of social conflicts.

The application to BiH considered the three groups to be homophilic. We note that if a group is relatively small (as in the Serbs' case) assuming that each individual interacts with everyone else is not unreasonable, and the mean-field results describe this well. If a group is large (e.g. Bosniaks), interactions may be more localized (among neighbors) with a distance decay effect on mutual persuasion. The Monte Carlo results represent this situation, which can yield chaotic time series of the attitudes. Results of chaotic patterns are unpredictable except for extremely short time ranges.

In general with asymmetric and opposite-sign Ks (like BiH) the temperature effects (corresponding to the intensity with which the context affects the conflict) are as follows: at low temperatures, each group sticks to its own stance in time, with little or no effect of the inter-group influences; at high temperatures, the system becomes disordered, and all attitudes average out to zero; at intermediate temperatures, attitudes move toward an attractor, constantly shifting among points on it but never settling on any. At the high end of intermediate temperatures, the attractor is continuous (i.e. there is an infinity of combinations of the average group attitudes); at the lower end of the  intermediate temperatures, the resulting  attractor is discrete, with a limited number of possible s combinations. The attractor morphology can be used to characterize the level of intractability, which increases with the temperature.

Interveners in conflicts can use the temperature results to strategize at a point in time (which is necessarily in a short time range, because the longer the time period considered, the more likely it is that parameters and context change): they may want to raise the temperature or lower it to defuse the conflict or even give it a push to agreement. However, altering the context (i.e. temperature) in a desired direction through intervention is a tall order. In reality, the kinds of actions intervenors can take correspond (in our model) to altering the J and K values in a conflict. Intractable conflicts require repeated interventions in time to change the system's path. 
Arguably, less external attention to, or pressure on the dynamics preceding the BiH elections might lower the temperature, but that is not necessarily in the interest of all the external players.

We note that for our choice of parameter values based loosely on the Prism survey, the BiH conflict appears intractable in the long run at all temperatures (Figures 2 to 5). Interveners and contextual events may alter the parameter values sufficiently to set it on a different course. Although additional analysis might surface the parameter values that might eventually yield settlement, it does not necessarily mean that we can attain them now through the right moves, since they result from a complex combination of future changes in parties and in their context.


Negotiation theory holds that qualitative differences between two and three parties in conflict exist due in part to an increase in process intricacies, to the possibility of coalitions and to the complexity of multiparty conflicts. Neither process nor coalition formation are directly captured by our model. Therefore, the two- and three-party models yield some qualitatively similar results. On the other hand, the increased level of complexity is reflected in the three-party model.

In the negotiation literature, some conflicts are discussed in terms of two main groups. Examples include Democrats and Republicans in the US, North Korea and South Korea, or Greeks and Turks in Cyprus. However, more groups may be involved in each case and may affect outcomes in time, either directly or indirectly through the conflict context (conceptualized in our model as temperature). Ignoring such groups has practical consequences. We may fail to make sense of observed outcomes and even resort to labeling the two disputing parties irrational to account for discrepancies between the observations and theory. We may also offer wrong remedies. If the conflict dividing the Korean peninsula is framed as between only two parties, we relegate some key others such as China to the background (temperature), despite the fact that they may be actively driving the outcomes. Therefore, temperature captures external contextual effects, but it is useful to scrutinize them and identify components that constitute additional parties. At times, what seems to be a large temperature impact is due to the inclusion in the context of a group with large clout. In the BiH case, we might consider the European Union an influence group important enough to pull out of the temperature effect. In other words, the temperature should account for diffuse or ill-understood influences rather than actions of another major group. Although we described in detail a three-party version, the model can accommodate additional parties (at the computational cost of additional parameters).

Our model is tailored qualitatively to the BiH case through the choice of parameter values. It can be tailored to other cases in the same way. This is an advantage and an important contribution. First, theory and practice indicate that few social conflict cases are sufficiently comparable; specifics make a sizeable difference, and there are never enough similar cases to allow derivation of more than very broad generalities. Second, since the model produces case-specific results, it avoids the mechanistic trap feared by Bernstein et al. (2000)\cite{Bernstein)} for sociophysics applications to social sciences. Instead, given a current starting point, it produces specific possible trajectories (or scenarios) for the consideration of stakeholders and interveners. In specific cases, however, the longer the time span, the less likely it is for the model parameters to remain unchanged, precisely because the conflict is dynamic, embedded in other changing systems, and subject to changing temperatures. Therefore, it is necessary to conduct sensitivity analyses on the choice of parameter values,\cite{foot7} as well as reassess these values whenever data become available (for example, an opinion survey conducted along similar lines as Prism after the 2018 BiH elections).

\section{Concluding remarks}\label{Conclusions}
In this article we have expanded a two-group  model of social conflict dynamics (Diep et al 2017) \cite{Diep2017} to three groups. It shows how conflicts among any number of groups can be analyzed. The groups we considered are homophilic, with individuals interacting mostly or only with members of their own group. The strength of the intragroup interactions J characterizes a group's cohesion: larger J values correspond to tightly linked groups, while lower J values corresponds to loosely connected groups. The interactions among the conflicting groups are captured by parameters K. Individuals' stances in one group are affected by the average stance of individuals in the other groups.The matrix of inter-group interactions is not necessarily symmetrical:$K_{m,n} \neq K_{n,m}$. Unlike physics phenomena that obey Newton's third law, in the world of humans the magnitudes of action and reaction are not necessarily equal. The effect of the actions of group $n$ on group $m$ can differ from the effect of group $m$ on group $n$. When $K_{m,n}$ and $K_{n,m}$ have opposite signs there is a `frustration-like' effect. While group $n$ acts on group $m$ positively, eliciting a "tit-for-tat"� response in group $m$ (i.e., similar values of the corresponding attitudes), group $m$ may acts on group $n$ negatively. The latter responds in a "contrarian"� fashion. For example, if group $m'$s average attitude is more ideological (positive value of the average stance) group n may respond in a conciliatory fashion (negative value of the average stance).  However, social frustration effects are different from physical ones: the former can occur for only two conflicting groups if $K_{1,2} * K_{2,1} < 0$, while the latter is a geometric effect occurring for example on a triangular lattice for antiferromagnetic interactions.  Note that intra-group frustration may be present  in reality. To induce a level of such internal frustrations, we can introduce a number of individuals having negative interactions with their neighbors \cite{DiepFSS}. Such a realistic model will be investigated in the future.

The three-group dynamics is generated  by assuming that attitudes at time t+1 are influenced by the attitudes of the average stance of other groups at an earlier time t, where time is measured in lag units.
The influence of the context on the individuals and the groups is represented by a temperature T. At high temperatures the average attitudes converge to zero (disorder, as all attitudes occur and average out to zero) while at low temperature they converge to the ground state values (order). At intermediate temperature interesting dynamic phenomena emerge. In the space of the mean attitudes $(s_1, s_2, s_3)$ the mean-field model exhibits an attractor that is either continuous or composed of a discrete number of points. We associate this attractor with intractable conflicts, which recur over long time periods while the disputants' attitudes oscillate between a large or small number of values respectively.

We used Monte Carlo simulations of the same model to explore the role of the range of interactions. While in the mean-field model individuals interact with same intensity no matter how far apart they are geographically or socially, in the Monte Carlo simulations the intragroup interactions are only between individuals  near one another. Then we no longer observe oscillations, but chaotic behavior emerges. Hence the intractable conflict outcomes are unpredictable because of the chaotic nature of its dynamics.







This work was supported by the Institute of Advanced Studies (IAS) of the University of Cergy Pontoise under the Paris Seine Initiative for Excellence  ("Investissements d'Avenir" ANR-16-IDEX-0008).
We thank the staff of the IAS for their hospitality.

{}


\begin{thebibliography}{}
\bibitem{foot1} Several other disciplines, including sociology, social psychology and political science have also examined the causes and patterns of social conflict.


\bibitem{Oberschall} Oberschall A. (1978), Theories of social conflict, Annual review of sociology, 4(1), 291-315.

\bibitem{Valacher} Vallacher R. R., Coleman P. T., Nowak A., \& Bui-Wrzosinska L. (2011), Rethinking intractable conflict: The perspective of dynamical systems, in Conflict, Interdependence, and Justice (pp. 65-94), Springer, New York, NY.

\bibitem{foot2}Examples of conflicts that have lasted for long time periods - and then ended - include the 100-year war between England and France (1337 to 1453); the South Africa struggle against Apartheid (1948 to the early 1990s); and the fight to end slavery and racial laws in the United States (which straddled two centuries). While the conflicts lasted, those who lived through them surely experienced them as intractable.

\bibitem{Bercovitch} Bercovitch J., Characteristics Of Intractable Conflicts, in Beyond Intractability. Eds. Guy Burgess and Heidi Burgess, Conflict Information Consortium, University of Colorado, Boulder. Posted: October 2003 $<$http://www.beyondintractability.org/essay/characteristics-ic$>$.

\bibitem{Burgess} Burgess H. and Burgess G. M., What Are Intractable Conflicts?,  in Beyond Intractability. Eds. Guy Burgess and Heidi Burgess. Conflict Information Consortium, University of Colorado, Boulder. Posted: November 2003 $<$http://www.beyondintractability.org/essay/meaning-intractability$>$.

\bibitem{Menczer} Menczer F. (2018), The spread of misinformation in social media,  Keynote address, NetSci Conference.

\bibitem{Bernstein} Bernstein S., Lebow R. N., Stein J. G., \& Weber S. (2000),  God gave physics the easy problems: adapting social science to an unpredictable world, European Journal of International Relations, 6(1), 43-76.

\bibitem{Kaufman1} Kaufman S. \& Kaufman M. (2013), Tipping points in the dynamics of peace and war, in International negotiation: Foundations, models and philosophies, 251-272.

\bibitem{Kaufman2} Kaufman S., \& Kaufman M. (2015), Two-Group Dynamic Conflict Scenarios: "Toy Model" with a Severity Index, Negotiation and Conflict Management Research, 8(1), 41-55.

\bibitem{Diep2017} Diep H. T., Kaufman M., \& Kaufman S. (2017), Dynamics of two-group conflicts: A statistical physics model. Physica A: Statistical Mechanics and its Applications 469, 183-199.

\bibitem{Kaufman3} Kaufman, M, Kaufman, S. \& Diep, H. T. (2017). Scenarios of Social Conflict Dynamics  on Duplex Networks.Journal on Policy and Complex Systems, 3(2), 3-13.

\bibitem{Wilson} Wilson A. (1969), Notes on some concepts in social physics, Papers in Regional Science 22(1), 159-193.

\bibitem{Stauffer} Stauffer D. (2003), Sociophysics simulations, Computing in Science \& Engineering 5(3), 71-75.

\bibitem{Barnes} Barnes T. J., \& Wilson M. W. (2014), Big data, social physics, and spatial analysis: The early years,  Big Data \& Society, 1(1), 2053951714535365.

\bibitem{Galam1}Galam S. (2012), Sociophysics: a physicist's modeling of psycho-political phenomena, Springer Science \& Business Media.


\bibitem{Godoy} Godoy A., Tabacof P., Von Zuben F.J. (2017), The role of the interaction network in the emergence of diversity of behavior, PLoS ONE 12(2): e0172073. doi:10.1371/journal.pone.0172073.

\bibitem{Schweitzer} Schweitzer F. (2018), Sociophysics, Physics Today 71, 2, 40

\bibitem{Galam2} Galam S. and Moscovici S. (1991), Towards a theory of collective phenomena: Consensus and attitude changes in groups, European Journal of Social Psychology 21: 49-74.

\bibitem{Stanley} Stanley H. E., Amaral L. A. N., Canning D., Gopikrishnan P., Lee Y., \& Liu Y. (1999), Econophysics: Can physicists contribute to the science of economics? Physica A: Statistical Mechanics and its Applications 269(1), 156-169.

\bibitem{Liben} Liben-Nowell D., \& Kleinberg J. (2007), The link-prediction problem for social networks, Journal of the Association for Information Science and Technology 58(7), 1019-1031.

\bibitem{Castellano} Castellano C., Fortunato S., \& Loreto V. (2009), Statistical physics of social dynamics, Reviews of modern physics 81(2), 591.
With this concept of universality in mind one can then approach the modelization of social systems, trying to include only the simplest and most important properties of single individuals and looking for qualitative features exhibited by models. A crucial step in this perspective is the comparison with empirical data which should be primarily intended as an investigation on whether the trends seen in real data are compatible with plausible microscopic modeling of the individuals, are self-consistent or require additional ingredients.

\bibitem{Helbing} Helbing D. (2010), Quantitative sociodynamics: stochastic methods and models of social interaction processes, Springer Science \& Business Media.

\bibitem{Coleman1} Coleman P. T., Vallacher R. R., Nowak A., \& Bui-Wrzosinska L. (2007), Intractable conflict as an attractor: A dynamical systems approach to conflict escalation and intractability, American Behavioral Scientist 50(11), 1454-1475.

\bibitem{Liebovitch} Liebovitch L. S., Naudot V., Vallacher R., Nowak A., Bui-Wrzosinska L., \& Coleman P. (2008), Dynamics of two-actor cooperation-competition conflict models. Physica A: Statistical Mechanics and its Applications 387(25), 6360-6378.

\bibitem{Majorana} Majorana E., \& Mantegna R. N. (2006), The value of statistical laws in physics and social sciences, in Ettore Majorana Scientific Papers (pp. 237-260), Springer, Berlin, Heidelberg.

\bibitem{Stewart} Stewart J. Q. (1950), The development of social physics. American Journal of Physics 18(5), 239-253.


\bibitem{Buchanan} Buchanan M. (2007), The social atom, Bloomsbury, New York, USA.

\bibitem{foot3} We argue in this article that in cases such as the ones we study, the symmetries are not preserved.


\bibitem{McPherson} McPherson M., Smith-Lovin L., and Cook J. M. (2001), Birds of a feather: Homophily in social networks, Annual review of sociology, 27(1), 415-444.

\bibitem{Aiello} Aiello L. M., Barrat A., Schifanella R., Cattuto C., Markines B., and Menczer F. (2012), Friendship prediction and homophily in social media, ACM Transactions on the Web (TWEB), 6(2), 9.

\bibitem{Diep2015} H. T. Diep, Statistical Physics: Fundamentals and Application to Condensed Matter, World Scientific (2015).

\bibitem{foot4} One example is the 2018 interaction North Korea - United States interaction. The US adopted a threatening stance at the outset; North Korea came to the negotiation table. As talks proceeded, however, and the US showed willingness to make some concessions, North Korea responded first by trying to withdraw from talks and later by adopting a more aggressive stance. On July 7, 2018, the US representative described recent talks as productive, while the North Korean representative called the same exchanges as regrettable and “gangster-like.

\bibitem{Tamkin} Tamkin E. (2018), Bosnia is teetering on the precipice of a political crisis, Foreign Policy https://foreignpolicy.com/2018/03/21/bosnia-is-teetering-on
    -the-precipice-of-a-political-crisis-balkans-election-law-dodik/.

\bibitem{Sito-Sucic} Sito-Sucic D. (2018), Ethnically tinged row over voting rules threatens governance in Bosnia, Reuters, February 23. https://www.reuters.com/article/us-bosnia-election-law/ethnically
    -tinged-row-over-voting-rules-threatens-governance-in-bosnia-idUSKCN1G72DC.

\bibitem{Toe} Toe R. (2016), Census Reveals Bosnia's Changed Demography, Balkan Insight, http://www.balkaninsight.com/en/article/new-demographic-picture-of-bosnia-finally-revealed-06-30-2016.

\bibitem{Mertus} Mertus J. (1997), Prospects for National Minorities under the Dayton Accords-Lessons from History: The Inter-War Minorities Schemes and the Yugoslav Nations, Brook J. Int'l L., 23, 793.

\bibitem{Friedman} Friedman F. (2013), Bosnia and Herzegovina: A polity on the brink. Routledge.



\bibitem{Szasz} Szasz P. C. (1997), The Dayton Accord: The Balkan Peace Agreement. Cornell Int'l LJ, 30, 759.

\bibitem{Prism} Prism Research (2013). Public Opinion Poll Results, Analitical Report. Office of the Un Resident Coordinator in Bosnia and Herzegovina.

\bibitem{foot5} Note that the survey was taken in 2013, about 5 years before the elections of 2018.

\bibitem{foot6} Tit-for-tat refers to a strategy used in the Prisoners' dilemma \cite{Axelrod} whereby at time $t$ a party responds to an opponent's move by doing exactly what the opponent did at time $t-1$.

\bibitem{Axelrod} Axelrod R., \& Hamilton W. D. (1981). The evolution of cooperation. science 211(4489), 1390-1396.

\bibitem{foot7} A sensitivity test entails examining ranges of $J$ and $K$ for which the model returns qualitatively similar results.

\bibitem{DiepFSS} Diep H. T. (Ed.), Frustrated Spin Systems, 2nd edition, World Scientific (2013).












%
%
%
%
%
%
%
%








\end{thebibliography}
\end{document}